\documentclass[11pt,a4paper]{article}
\usepackage{amsmath}
\usepackage{fancybox}
\usepackage{color}
\usepackage{pstricks}

\usepackage{young}
\usepackage{color}
\usepackage{pstricks}
\usepackage{graphicx}

\usepackage[T1]{fontenc}
\usepackage[utf8]{inputenc}
\usepackage{authblk}
\title{ D3-D5 Holography with   Flux}
\author[1]{Gianluca Grignani}
\author[2]{Namshik Kim}
\author[2]{Gordon W. Semenoff}
\affil[1]{\small Dipartimento di Fisica, Universit\`a di Perugia, I.N.F.N. Sezione di Perugia, Via Pascoli, I-06123 Perugia, Italy}
\affil[2]{   Department of Physics and Astronomy, University of British Columbia, 6224 Agricultural Road, 
Vancouver, British Columbia, Canada V6T 1Z1}

\date{}

\begin{document}

  \maketitle

\abstract{It is shown that the Berezinski-Kosterlitz-Thouless phase transition that
has been found in D3-D5 brane systems with nonzero magnetic field and charge density can
also be found by tuning an extra-dimensional magnetic flux. We find numerical solutions
for the probe D5-brane embedding and discuss properties of the
solutions. We also demonstrate that the nontrivial embeddings include
those which can be regarded as spontaneously breaking chiral symmetry.}

\vskip 2cm



The AdS/CFT duality of an appropriately oriented probe D5-brane embedded
in $AdS_5\times S^5$ space-time
and a supersymmetric defect conformal field theory
is a well-studied example of holography \cite{Karch:2000gx}-\cite{Evans:2010hi}.
In the limit of large $N$ and large radius of curvature,   the D5-brane geometry is found as
an extremum of the Dirac-Born-Infeld action with appropriate Wess-Zumino terms added.
Its world-volume is the product space
$AdS_4(\subset AdS_5)\times S^2(\subset S^5)$ which preserves an $OSp(4|4)$ subgroup of the $SU(2,2|4)$
superconformal symmetry of the $AdS_5\times S^5$ background.
The superconformal  field theory which is dual to this D3-D5 system,
and which is described by it in the strong coupling limit,
has a  co-dimension one membrane that is embedded in $3+1$-dimensional flat space.
The bulk of the 3+1-dimensional  space is occupied by
${\mathcal N}=4$ supersymmetric Yang-Mills theory with $SU(N)$ gauge group.    A bi-fundamental chiral hypermultiplet
lives on the membrane defect and its field theory is dual to the low energy
modes of open strings connecting the D5-branes and the D3-branes.  These fields transform in the fundamental representation
of the $SU(N)$ bulk gauge group and in the fundamental representation of the global $U(N_5)$, where $N_5$ is the number
of D5-branes (in the probe limit,  $N_5<<N$ and  we will take $N_5=1$).    The defect field theory preserves half of the supersymmetries of the bulk ${\mathcal N}=4$ theory, resulting in
the residual $OSp(4|4)$ super-conformal symmetry.  It is massless with a hypermultiplet
mass operator which breaks an  SU(2) R-symmetry \cite{DeWolfe:2001pq}.

An external magnetic field has a profound effect on this system.
In the quantum field theory, the magnetic field is constant and is perpendicular
to the membrane defect.  In the string theory,  the magnetic field
destabilizes the conformal symmetric state to one which spontaneously breaks the SU(2)
R symmetry and generates a mass gap for the D3-D5 strings \cite{Filev:2009xp}.
The only solution for the D5-brane embedding has it pinching
 off before it reaches the Poincar\'e horizon of $AdS_5$.  As a result, the D3-D5 strings which, when excited, must
 reach from the D5-brane to the  Poincar\'e horizon, have a minimum length and an energy gap.
 This occurs for any value of the magnetic field, in fact, since the theory has conformal invariance,
the magnetic field is the only dimensional parameter and there is no distinction between large field and small field.
A mass and a mass operator condensate for the  D3-D5 strings can readily be identified 
(the  conformal dimensions of their field theory duals are
protected by supersymmetry) and there is simply no solution of the probe D5-brane embedding problem with a magnetic field
when both the mass and the condensate are zero.
There can be a solution when one of those parameters vanishes and the other does not vanish.
Such a solution can be interpreted as presence of a condensate in the absence
of a mass operator, that is, as dynamical symmetry breaking.
This phenomenon is regarded as a holographic realization of the ``magnetic catalysis'' of chiral symmetry breaking that has been
studied in 2+1-dimensional quantum field theories \cite{cat0}-\cite{tong}. The field theory studies rely on weak coupling expansions and re-summation of Feynman diagrams.  Whether the phenomenon can persist at strong
coupling is an interesting question which appears to have an affirmative answer in the context of this construction.
It and many other aspects of
the phase diagram of the D5-brane have been well studied in what is by now an extensive literature \cite{Filev:2009xp}-\cite{Evans:2010new}.

This interesting behavior becomes more complex when a U(1) charge density,  as well as the magnetic field, 
is introduced.  The state then
 has a non-zero density of D3-D5 strings. There is also a tuneable dimensionless parameter, the ratio of charge density to the field, the ``filling fraction''  $\nu=\frac{2\pi\rho}{B}$.
 In this case, there is no charge gap.  The D5-brane must necessarily reach the Poincar\'e horizon.
This is due to the fact that, to have a nonzero charge density,
there must be a density of fundamental strings suspended between the D5-brane and the Poincar\'e horizon.
 However, the fundamental string tension is always greater than the D5-brane tension \cite{Myers:2008me} and such strings  would therefore pull the D5-brane to the horizon.  The result is a gapless state:  the D3-D5 strings could have zero length, and therefore have no energy gap.  At weak coupling, the dual process is the formation of a fermi surface and a gapless metallic state when the charge density is nonzero.

What is more, if the filling fraction is large enough, the state with no mass term and   mass operator condensate equal to zero  exists and is stable. In this state, the SU(2) R symmetry is not broken.
As the filling fraction is lowered from large values where the system takes up this symmetric phase,
as pointed out in the beautiful paper \cite{Jensen:2010ga}, the system undergoes a Berezinski-Kosterlitz-Thouless-like (BKT) phase transition.  This phase transition has BKT scaling and is one of the rare examples on non-mean field phase transitions
in holographic systems.   When the filling fraction is less than the critical value, again, even though the D5-brane world-volume now reaches the Poincar\'e horizon, there is no solution of the theory unless either the mass operator or mass operator condensate or both are turned on. This state breaks the SU(2) R symmetry but still has no charge gap.

In this Letter, we shall observe that, as well as density, there is a second parameter which can drive the BKT-like
transition.  The parameter is the value of a magnetic flux which
 forms a U(1) monopole bundle on the D5-brane world-volume 2-sphere.
The possibility of adding this flux was suggested by Myers and Wapler \cite{Myers:2008me}.
They found that the idea could be used to construct stable
D3-D7 systems, in particular, and a modification of their idea was
subsequently used  to study holography in
D3-D7 systems \cite{Bergman:2010gm}-\cite{Davis:2011am}.
In the limit where the string theory is classical, the problem of embedding a D5-brane in the $AdS_5\times S^5$ geometry
reduces to that of finding an extremum of the Dirac-Born-Infeld and Wess-Zumino actions,
\begin{align}\label{dbi}
S=\frac{ T_5}{g_s} \int d^6\sigma\left[- \sqrt{-\det( g+2\pi\alpha' F)} + C^{(4)}\wedge2\pi\alpha' F\right] 
\end{align}
where $g_s$ is the closed string coupling constant, which is related to the ${\mathcal N}=4$
Yang-Mills coupling
by $4\pi g_s =g_{YM}^2$, $g_{ab}(\sigma)$ is the induced metric
of the D5 brane, $C^{(4)}$ is the  4-form of the $AdS_5\times S^5$ background,
$F$ is the world-volume gauge field and
$T_5=\frac{1}{(2\pi)^5{\alpha'}^3}$.
We shall use  the metric of $AdS_5\times S^5$ and 4-form
\begin{align}\label{ads5metric}
ds^2= L^2&\left[ r^2 (-dt^2+dx^2+dy^2+dx^2) + \frac{dr^2}{r^2}+\right. \nonumber \\
&\left. +d\psi^2+\cos^2\psi(d\theta^2+\sin^2\theta d\phi^2)+ \sin^2\psi (d\tilde\theta^2+\sin^2\tilde\theta d\tilde\phi^2)\right]
\\ \label{4form}
C^{(4)}=  L^4&r^4dt\wedge dx\wedge dy\wedge dz+L^4\frac{c(\psi)}{2}d\cos\theta\wedge d\phi
\wedge d\cos\tilde\theta\wedge d\tilde\phi
\end{align}
with
$
\partial_\psi c(\psi)=8\sin^2\psi\cos^2\psi
$.
In (\ref{ads5metric}), the 5-sphere is represented by two 2-spheres fibered over the interval $\psi\in[0,\tfrac{\pi}{2}]$.
The radius of curvature of $AdS$ is $L$ and  $L^2=\sqrt{\lambda}\alpha'$ with $\lambda=g_{YM}^2N$.
The embedding of the D5-brane is mostly determined by symmetry.  The dynamical variables are
$
\{ x(\sigma),y(\sigma),z(\sigma),t(\sigma),r(\sigma),\psi(\sigma),\theta(\sigma),\phi(\sigma),
\tilde\theta(\sigma),\tilde\phi(\sigma)\}
.$  We look for a solution of the form
\begin{align}\label{ansatz}
\sigma_1=x,\sigma_2=y,\sigma_3=t,\sigma_4=r,
\sigma_5=\theta-\tfrac{\pi}{2},\sigma_6=\phi,\tilde\theta=0,\tilde\phi=0
\end{align} and  the remaining coordinates depending only
on $\sigma_4=r$, ($z(r), \psi(r)$).\footnote{  This ansatz is symmetric under spacetime parity which can
be defined for the Wess-Zumino terms
\begin{align}
\label{cs1}
& \int d^6\sigma \epsilon^{\mu_1\mu_2\ldots\mu_6} \partial_{\mu_1}x(\sigma)\partial_{\mu_2}y(\sigma) \partial_{\mu_3} z(\sigma)\partial_{\mu_4}t(\sigma)  r^4(\sigma) ~\partial_{\mu_5}A_{\mu_6}(\sigma)
\\ \label{cs2}
& \int d^6\sigma \epsilon^{\mu_1\mu_2\ldots\mu_6} \partial_{\mu_1}\cos\theta(\sigma)\partial_{\mu_2}\phi(\sigma)
 ~\partial_{\mu_3} \cos\tilde\theta(\sigma)\partial_{\mu_4}\tilde\phi(\sigma)
 c(\psi)  \partial_{\mu_5} ~A_{\mu_6}(\sigma)
\end{align}
in the following way.  The world-volume coordinates transform as
$
\left\{\sigma_1',\sigma_2',\ldots,\sigma_6'\right\}=\left\{-\sigma_1,\sigma_2,\ldots,\sigma_6\right\}
$
and the embedding functions as  $x'(\sigma')=-x(\sigma)$, $ \tilde\theta'(\sigma')=\pi-\theta(\sigma)$,
$A_1'(\sigma') = -A_1(\sigma)$ with all other variables obeying $\chi(\sigma')=\chi(\sigma)$.
This is a symmetry of the Wess-Zumino terms and the Ans\"atz (\ref{ansatz}) is invariant.
Charge conjugation flips the sign of all gauge fields, $A\to -A$ and we augment it by $
\left\{\sigma_1',\ldots,\sigma_5',\sigma_6'\right\}=\left\{\sigma_1,\ldots,-\sigma_5,\sigma_6\right\}
$.  The Wess-Zumino terms are invariant.  The background field $fd\cos\theta\wedge d\phi$ is also invariant
once we choose $\sigma_5=\theta-\frac{\pi}{2}$. The fields $a(r)$ breaks C and preserves P.  $b$ breaks C and P and preserves
CP.}
With this Ans\"atz, the D5-brane world-volume metric is
\begin{align}\label{d5metric}
ds^2= L^2\left[ r^2 (-dt^2+dx^2+dy^2)+\frac{dr^2}{r^2}(1+r^2{\psi'}^2+r^4{z'}^2)+ \cos^2\psi(d\theta^2+\sin^2\theta d\phi^2) \right]
\end{align}
where prime denotes derivative by $r$
and the world-volume gauge fields are
\begin{align}\label{magneticfields}
F= \frac{L^2}{2\pi\alpha'}~  a'(r)~dr\wedge dt+  \frac{L^2}{2\pi\alpha'}~b~dx\wedge dy+ \frac{L^2}{2\pi\alpha'}~\frac{f}{2}~d\cos\theta\wedge d\phi
\end{align}
Here, $f$ is the strength of the monopole bundle.\footnote{A monopole bundle has quantized flux.  Here the number of quanta
  is very large in the strong coupling limit $n_D\sim\sqrt{\lambda}$, so that it is to a good approximation a continuously
  variable parameter.  $b$ and $q$ are related to the physical magnetic field and charge density as
$
b=\frac{2\pi}{\sqrt{\lambda}}B$, $q= \frac{4\pi^3}{\sqrt{\lambda}N }\rho
$
so that
$
\frac{q}{b}= \frac{\pi}{N }\frac{2\pi\rho}{B}\equiv\frac{\pi}{N }\nu
$
where the dimensionless parameter $\nu$ is the filling fraction. A Landau level would have degeneracy $N$. The filling
fraction of a set of N degenerate levels naturally scales like $N$ to give order one $b$ and $q$ in the large $N$ limit.}   $b$ is a constant magnetic field
which is proportional to a constant magnetic field in the field theory dual. $a(r)$ is the temporal
world-volume gauge field which must be non-zero in order to have a uniform  charge density
in the field theory dual.  The bosonic part of the R symmetry is $SU(2)\times SU(2)$. One SU(2) is the isometry of the $S^2$ which is wrapped by the D5 brane (\ref{d5metric}) and is also a symmetry of the background fields (\ref{magneticfields}).  The other is the rotation in the transverse $S^2\subset S^5$ with $S^5$ coordinates $\tilde\theta,\tilde\phi$. This is a symmetry of the embedding  only when the former $S^2$ is maximal, that is, when $\psi(r)=0$ for all $r$. If $\psi(r)$
deviates from zero, it must choose a direction in the transverse space, and the choice breaks the second SU(2).
The hypermultiplet mass shows up in the D5 brane embedding as
\begin{align}\label{hypermultipletmass}
M\sim m\equiv\lim_{r\to\infty}r\sin\psi(r)~~,~~\psi(r\to\infty)=\frac{m}{r}+\frac{c}{r^2}+\ldots
\end{align}
and deviation of $\psi(r)$
from the constant $\psi=0$ so that the parameter $m$ is nonzero is a signal of having switched on a hypermultiplet
mass operator in the
dual field theory. The parameter $c$ is dual to the chiral condensate, although in an alternative quantization these could
be interchanged \cite{Klebanov:1999tb}.

With (\ref{d5metric}) and (\ref{magneticfields}), the action (\ref{dbi}) is
\begin{align}\label{ansatzaction}
S={\mathcal N}\int d^3xdr\left[- \sqrt{(f^2+4\cos^4\psi)(b^2+r^4)(1+r^2{\psi'}^2+r^4{z'}^2)-{a'}^2}+    f r^4z'\right]
\end{align}
where
$
{\mathcal N}=\frac{2\pi T_5 L^6 }{g_s}=\frac{\sqrt{\lambda} N}{4\pi^3}
$.
The factor of $2\pi$ in the numerator comes from half of the volume of the unit 2-sphere (the other factor of 2 is still in the action).
The Wess-Zumino term gives a source for $z(r)$.

Now, we must solve the equations of motion for the functions $\psi(r)$, $a(r)$ and $z(r)$ which result
from (\ref{ansatzaction}) and the variational principle. The variables  $a(r)$ and $z(r)$
are cyclic and they can be eliminated using their equations of motion,
\begin{align}
&\frac{d}{dr}\frac{\delta S}{\delta z'(r)}=0~~\to~~  \frac{\sqrt{(f^2+4\cos^4\psi)(b^2+r^4)}r^4z'}{\sqrt{1+r^2{\psi'}^2+r^4{z'}^2-{a'}^2}}-    f r^4 =p_z\\
&\frac{d}{dr}\frac{\delta S}{\delta a'(r)}=0~~\to~~
  \frac{\sqrt{(f^2+4\cos^4\psi)(b^2+r^4)} {a'} }{\sqrt{1+r^2{\psi'}^2+r^4{z'}^2-{a'}^2}}=-q
\end{align}
where $ p_z$ and $q$ are constants of integration.
If these equations are to hold near $r\to0$, we must set $p_z=0$. $q$ is proportional to the charge density
in the field theory dual. Then, we
can solve for $z'$ and $a'$,
\begin{align}
\label{equationforzp}
&z'=\frac{ f\sqrt{1+r^2{\psi'}^2}}{\sqrt{ 4\cos^4\psi (b^2+r^4)+f^2b^2+q^2}}
\\
&a'=\frac{-q\sqrt{1+r^2{\psi'}^2}}{\sqrt{ 4\cos^4\psi (b^2+r^4)+f^2b^2+q^2}} \label{aprime}
\end{align}
We must then use the Legendre transformation
$$
{\mathcal R}= S-\int a'(r)\frac{\partial L}{\partial a'(r)}-\int z'(r)\frac{\partial L}{\partial z'(r)}
$$
to eliminate $z'$ and $a'$.  We obtain the Routhian
\begin{align}
{\mathcal R}=-{\mathcal N} \int d^3xd r \sqrt{4\cos^4\psi(b^2+r^4)+ b^2f^2+q^2}\sqrt{1+r^2{\psi'}^2}
\label{routhian1}
\end{align}
which must now be used to find an equation of motion for  $\psi(r)$,
\begin{align}
\frac{\ddot\psi}{1+\dot\psi^2}+\dot\psi\left[ 1+\frac{8r^4\cos^4\psi}{4(b^2+r^4)\cos^4\psi+f^2b^2+q^2}\right] +
\frac{8(b^2+r^4)\cos^3\psi\sin\psi}{4(b^2+r^4)\cos^4\psi+f^2b^2+q^2}=0
\label{equationforpsi}\end{align}
where the overdot is the logarithmic derivative  $\dot\psi=r\frac{d}{dr}\psi$.

First, we note that, if $\psi(r)$ is to be finite at $r\to\infty$, its logarithmic derivatives should vanish.  Then, the only
boundary condition which is compatible with the equation of motion is $\psi(r\to\infty)=0$. The asymptotic solution of
(\ref{equationforpsi}) at large $r$ is given in (\ref{hypermultipletmass}).

If we set $b=0$, $f$ does not appear in the Routhian (\ref{routhian1}) or in the equation of motion (\ref{equationforpsi}).
$\psi(r)$ which is then $f$-independent.
In fact, when $b=0$, the constant solution $\psi=0$ is a stable solution of (\ref{equationforpsi}).   $z(r)$ is   $f$ and $r$-dependent.
Equation
(\ref{equationforzp}) has the solution
$
z(r)=\int  d r\frac{ f }{\sqrt{ 4   r^4 +f^2 }} = r \, _2F_1\left(\frac{1}{4},\frac{1}{2};\frac{5}{4};-\frac{4 r^4}{f^2}\right)
$.
The worldvolume is  $AdS_4\times S^2$,
where the radii of the two spaces differ, the $S^2$  has radius $L$ whereas $AdS_4$ has
radius $L\sqrt{1+\tfrac{f^2}{4}}$. The field theory dual of this system was discussed in reference \cite{Myers:2008me}.
It  has a planar defect dividing three dimensional space into two half-spaces
with ${\mathcal N}=4$ Yang-Mills theory with gauge group  $SU(N+n_D)$
on one side of the defect and ${\mathcal N}=4$ Yang-Mills theory with gauge group $SU(N )$
on the other side.  Here $n_D$ is the number
of Dirac monopole quanta in $f$.   The $r$-dependence of the embedding function $z(r)$ can
be viewed as an energy-scale dependent position of the defect in the field theory.

\begin{figure}\includegraphics[scale=.7]{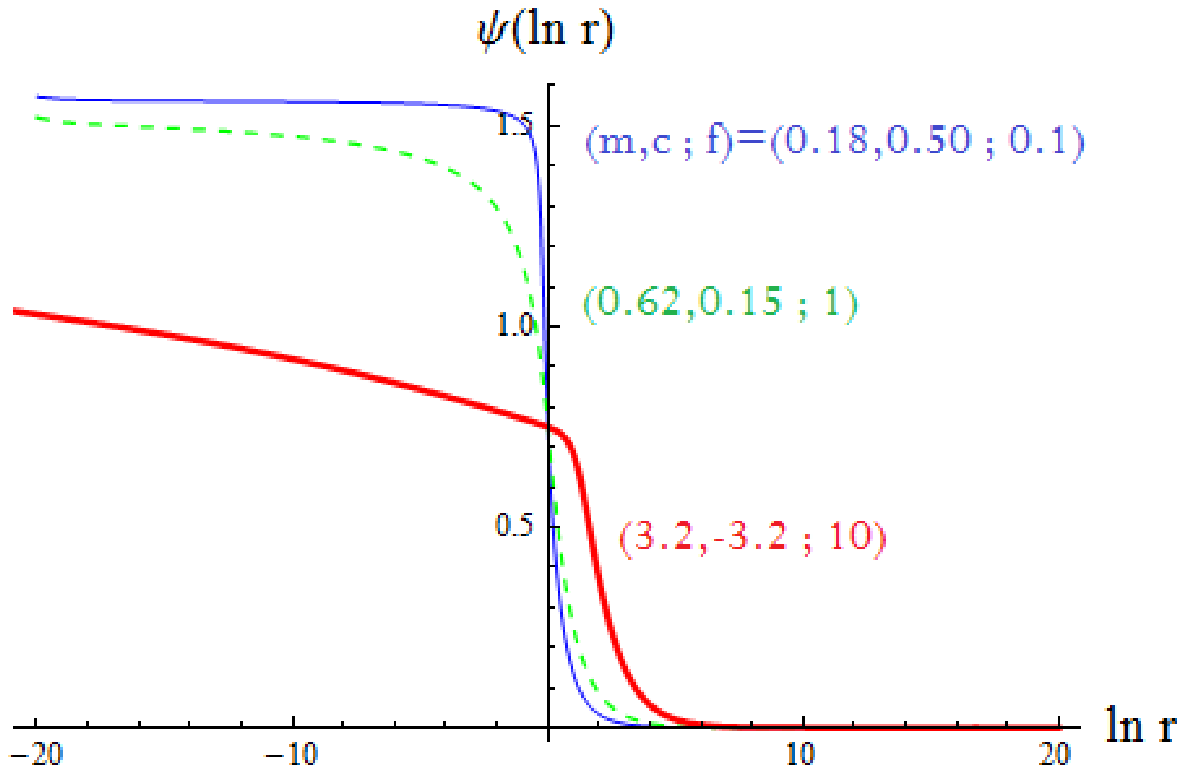}
\begin{caption}
{Equation (\ref{equationforpsi}) integrated with $q=0$ and $f^2=0.01$
 $f^2=1$   and   $f^2=100$.    The solutions
interpolate  between the correct asymptotic values, $\psi(r=\infty)=0$ and $\psi(r=0)=\frac{\pi}{2}$.
The AdS radius r is measured in units of $1/\sqrt{b}$.  For smaller values of $f$, the transition is sharp
and occurs at $ \sqrt{b}~r\sim1$. \label{psi0}
}\end{caption}
\end{figure}

When $b$ is not zero, scaling $r\to\sqrt{b}~r$,
removes $b$ from most of equation (\ref{equationforpsi}), the dependence which remains is only
in the parameter $f^2+\left(\tfrac{q}{b}\right)^2$. If this parameter is large enough, the solution $\psi(r)=0$
is still a stable solution of (\ref{equationforpsi}). When $f^2+\left(\tfrac{q}{b}\right)^2$ is lowered to a
critical value, the $\psi=0$ solution
becomes unstable. At that point, the BKT-like phase transition occurs.
That transition was found in reference \cite{Jensen:2010ga} where they adjusted $\tfrac{q}{b}$ (they had $f=0$) with the critical value being $ \left.\left(\tfrac{q}{b}\right)^2\right|_{\rm crit.}=28$.
The onset of instability of the symmetric solution $\psi =0$
 at that point is easily seen by looking at solutions
 of the linearized equation which, at small $r$, must be $\psi\sim c_1r^{\nu_+}+
 c_2r^{\nu_-} $ with $\nu_\pm=-\frac{1}{2}\pm\frac{1}{2}\sqrt{ 1-32/\left(4+f^2+\tfrac{q}{b}\right)^2 }
$.  The instability sets in when the exponents become complex, that is, at $\left[f^2+\left(\tfrac{q}{b}\right)^2\right]_{\rm crit.}= {28}$.  The complex exponents are due to the fact that, in the $r\sim 0$ regime, the
fluctuations obey a wave equation for $AdS_2$ with a mass that violates the Breitenholder-Freedman bound.
Since, in the stable regime, $f^2+\left(\tfrac{q}{b}\right)^2> {28}$ both of the exponents in the fluctuations are negative, deviation
from $\psi(r)=0$ is not allowed, it is an isolated solution.
Here we point out that we can find   the phase transition  even when $\tfrac{q}{b}$ vanishes
by varying $f$, stability occurs where $f^2>28$
and the phase transition at $f^2_{\rm crit}=28$. In particular,
 this allows us to study the theory in the charge neutral state where $q=0$.

\begin{figure}
 \includegraphics[scale=.6]{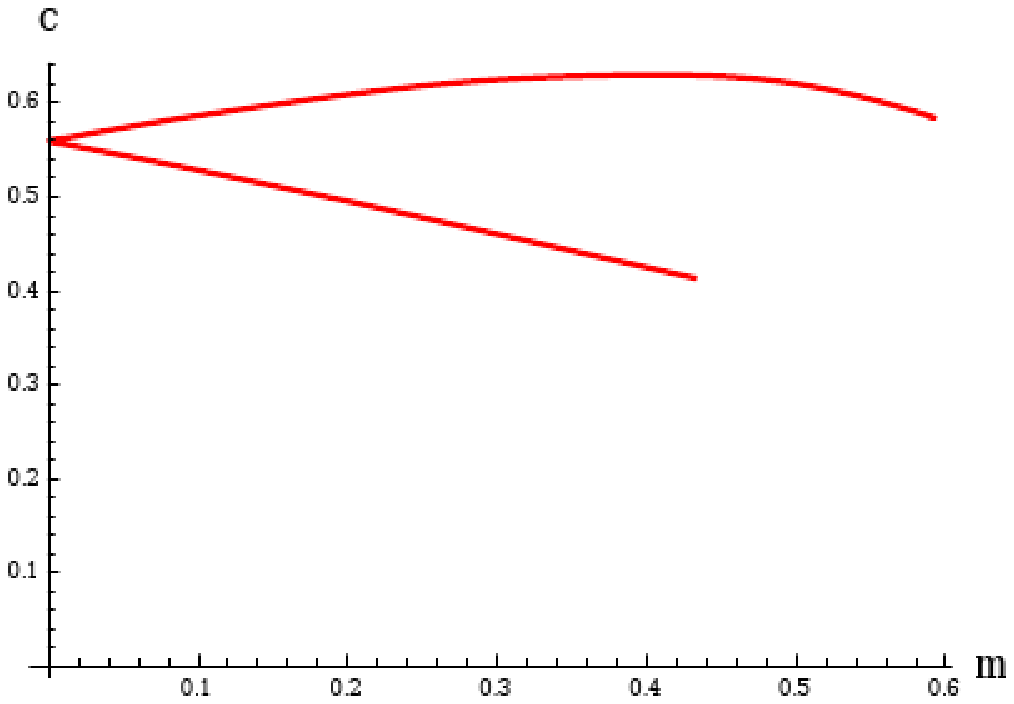}\\
\begin{caption} {The constants $c$ versus $m$ are plotted for a sequence of  embeddings of the D5-brane in the region where the constant $\psi$ solutions
are unstable,  $f^2=.01$.
\label{cvsm1}
}\end{caption}
\end{figure}

When the symmetric solution $\psi =0$ is unstable, we must find another solution of  equation (\ref{equationforpsi}) for $\psi(r)$,
where we now assume that it depends on $r$.  $\psi=0$ was an isolated solution, there are no other solutions closeby.
As soon as it depends on $r$, if $\psi(r)$ is to remain finite in the small r region, it must go to the other solution of (\ref{equationforpsi}) at small $r$, $\psi(r\to0)=\tfrac{\pi}{2}$. At this point, the $S^2$ which the D5-brane wraps has
collapsed to a point and the D5-brane is effectively a D3-brane with worldvolume oriented in the $x,y,t,r$-directions.

\begin{figure}
 \includegraphics[scale=.7]{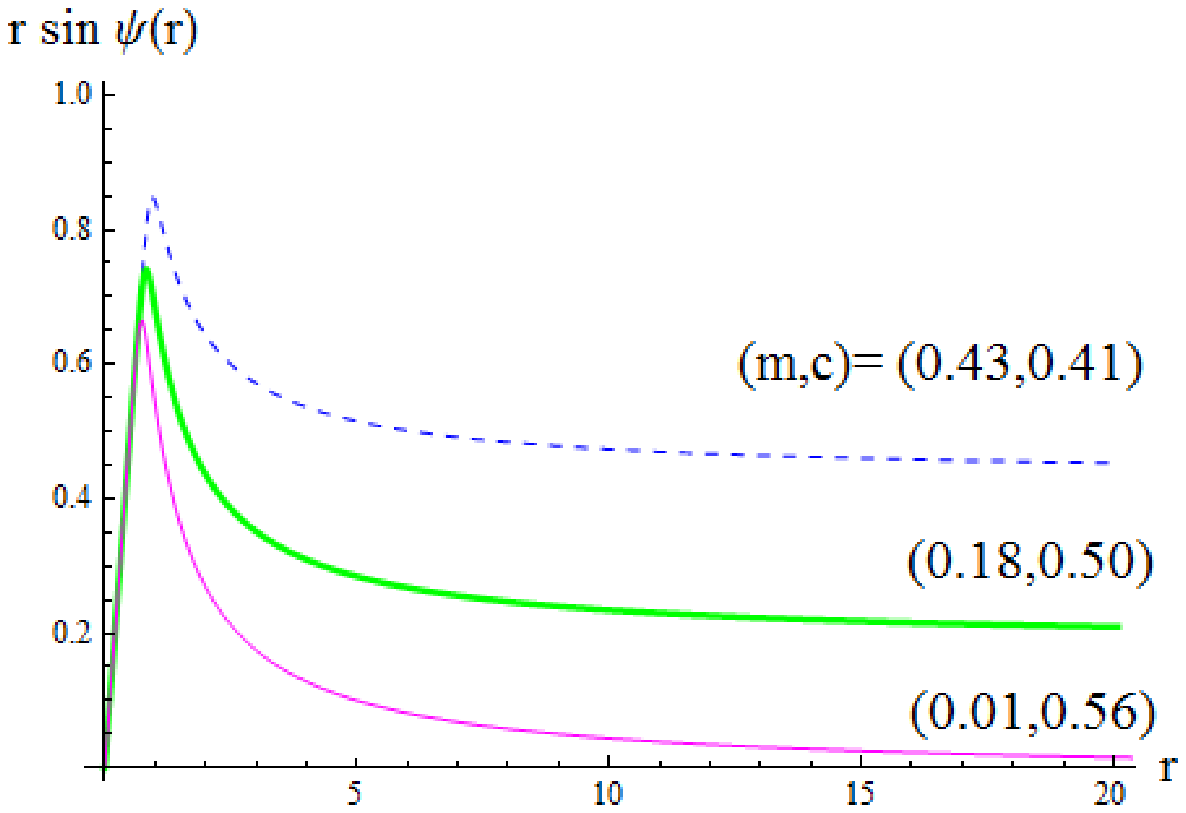}\\
\begin{caption} {The function $r\sin(\psi(r))$ is plotted versus $r$ for some embeddings, parameterized
by the asymptotic $m$ and $c$, including the one which is close the solution with $m=0$ which is associated with dynamical symmetry breaking.
\label{rsinpsi}
}\end{caption}
\end{figure}

When $q$, $b$ and $f$ are all nonzero, it is interesting that
the embedding problem depends only on the combination $\sqrt{f^2+\left(\tfrac{q}{b}\right)^2}$, reminiscent of
bound states of F-strings and D-branes \cite{Witten:1995im}.
When either or both of $q$ and $f$ are nonzero, the D5-brane must reach the Poincar\'e horizon.
Otherwise, the charge density $q$ and magnetic monopole flux $f$ would have to have  sources on the D5-brane worldvolume.
The appropriate source would be $n_D$ D3-branes carrying electric charge density $q$  suspended between the
D5 world-volume and the Poincar\'e horizon.   However, as in the case of fundamental strings when there was only charge
present,
it is possible to show that the appropriate D3-brane tension is always greater than the D5-brane tension.  The suspended D3-branes would
 pull the D5-brane to the horizon.  The D5-brane world-volume could still reflect this behavior with a spike or funnel-like configuration which emulates suspended strings and D3-branes in the $r<1/\sqrt{b}$ regime.

 We  expect to find solutions of (\ref{equationforpsi}) which interpolate between $\psi=0$ at $r\to\infty$ to $\psi=\tfrac{\pi}{2}$ at $r\to 0$. Indeed, for generic asymptotic behavior, such solutions are easy to
find by a shooting technique.  Examples are given in figure \ref{psi0}. Indeed we see that, as a function of $\ln(\sqrt{b}r)$
and when $f^2+\left(\tfrac{q}{b}\right)^2<\left[f^2+\left(\tfrac{q}{b}\right)^2\right]_{\rm crit.}^2=28$, they exhibit a rapid soliton-like crossover between $\psi=0$ at large $\sqrt{b}r$, which is a D5-brane,   and $\psi\sim \tfrac{\pi}{2}$ at small $\sqrt{b}r$, which is like $n_D$ D3-branes with
electric charge  $q$ dissolved into them. In the third plot in figure \ref{psi0}, where $f^2+\left(\tfrac{q}{b}\right)^2>\left[f^2+\left(\tfrac{q}{b}\right)^2\right]_{\rm crit.}^2$, the
funnel is much more diffuse.

It is also possible to find solutions that can be interpreted as chiral symmetry breaking, although the D5-brane still reaches
the Poincar\'e horizon and we expect that the D3-D5 strings are still gapless.  In the region of large $r$, the linearized equation for $\psi(r)$ is solved by (\ref{hypermultipletmass}).
The two asymptotic behaviors have power laws associated with the ultraviolet conformal dimensions of the
mass  and the chiral condensate in the dual field theory. These are the same as their classical dimensions since they are
protected by supersymmetry. A symmetry breaking solution would have one of these equal to zero (and the other
one interpreted as a condensate).  Indeed, it is easy to find a family of solutions of (\ref{equationforpsi})
which, as we tune $m$, still exists and has nonzero $c$ in the limit where $m$ goes to zero.  The $c$ versus
$m$ behavior of this family of solutions is shown in figure \ref{cvsm1}. The behavior if $r\sin(\psi(r))$ which
can be interpreted as the separation of the D5 and D3-branes is plotted in figure \ref{rsinpsi} for some values
of $m$ and $c$.

As an extension of our results, it would be interesting to analyze the electromagnetic properties of
the solution with finite $f$ and $q=0$.  This is a charge neutral state and it has a mass operator condensate. It is possible to study Maxwell's equations for fluctuations of the world-volume gauge field and though it is difficult to fully derive even a formal solution, it is relatively straightforward to show that they have no solution when the field strength is a constant.  This implies that the charged matter is still gapless and provides the singularities in response functions which make the theory singular at low energy and momentum.

From the point of view of the space-time symmetry, the flux $f$ is charge conjugation symmetric, whereas the
finite charge density state is not.  In fact $f$ itself does not violate any 2+1-dimensional spacetime symmetries associated with Lorentz, C, P or T invariance.  The fact that states with finite $f$ need to have a vanishing charge gap is somewhat
mysterious from the field theory point of view.
When $q$ is nonzero, the absence of a charge gap is understandable as the
 theory should be in a finite density metallic state, even when it breaks chiral symmetry.  When $q=0$ but $f$ is nonzero, the system must also be gapless, even though the charge density is zero and the hypermultiplet should be massive.
One possibility is that, in the field
theory, the planar defect which separates spaces where ${\cal N}=4$ Yang-Mills theory has different
gauge groups has a band of
gapless edge states. It would be interesting to examine this further in the field theory.

 \noindent
This work is supported in part by NSERC of Canada and in part by the MIUR-PRIN contract 2009-KHZKRX.

\end{document}